\documentclass{moriond}
\usepackage{amsfonts}
\usepackage[ansinew]{inputenc}      
\usepackage{amssymb}                
\usepackage{amsmath}                
\usepackage[]{graphicx}
\usepackage{floatflt}
\usepackage{mathrsfs}
\usepackage{verbatim}
\usepackage{overpic}
\usepackage{xspace}
\usepackage{lineno}
\usepackage{enumitem}
\usepackage[font=small,skip=3mm]{caption}
\usepackage{setspace}
\usepackage{array}
\usepackage{lscape}
\usepackage{rotating}

\def\be{\begin{equation}}
\def\ee{\end{equation}}
\def\bea{\begin{eqnarray}}
\def\eea{\end{eqnarray}}

\newcommand{\ttbar}{\ensuremath{t\bar{t}}\xspace}
\newcommand{\etmiss}{\ensuremath{E \kern-0.6em\slash_{\rm T}}\xspace}
\newcommand{\etmissx}{\ensuremath{E \kern-0.6em\slash_{\rm x}}\xspace}
\newcommand{\etmissy}{\ensuremath{E \kern-0.6em\slash_{\rm y}}\xspace}

\newcommand{\ljets}{\ensuremath{\ell\!+\!{\rm jets}}\xspace}

\newcommand{\etal}{\textit{et~al.}}

\newcommand{\GeV}{\ensuremath{\textnormal{GeV}}\xspace}
\newcommand{\TeV}{\ensuremath{\textnormal{TeV}}\xspace}

\newcommand{\dif}{\ensuremath{{\rm d}}}

\newcommand{\met}{\ensuremath{E\!\!\!\!/_T}\xspace}

\newcommand{\fb}{\ensuremath{{\rm fb}^{-1}}\xspace}

\newcommand{\mt}{\ensuremath{m_t}\xspace}

\newcommand{\pt}{\ensuremath{p_T}\xspace}

\newcommand{\stt}{\ensuremath{\sigma_{t\bar t}}\xspace}
\newcommand{\afb}{\ensuremath{A_{\rm FB}}\xspace}
\newcommand{\afbtt}{\ensuremath{A_{\rm FB}^{\ttbar}}\xspace}
\newcommand{\afbl}{\ensuremath{A_{\rm FB}^{\ell}}\xspace}

\newcommand{\Photo}{}

\begin{document}
\begin{flushright}
FERMILAB-CONF-15-097-E
\end{flushright}
\vspace*{4cm}
\title{RECENT TOP QUARK PRODUCTION RESULTS FROM THE TEVATRON}

\author{{\sc O. Brandt} on behalf of the {\sc CDF} and {\sc D0 Collaborations}}

\address{Kirchhoff-Institut f\"ur Physik, Im Neuenheimer Feld 227,\\
69120 Heidelberg, Germany}


\maketitle
\abstracts{In this article, I review recent measurements of the production of the top quark in $p\bar p$ collisions at a centre-of-mass energy of $\sqrt s=1.96~\TeV$ in Run II of the Fermilab Tevatron Collider, recorded by the CDF and D0 Collaborations. I will present the Tevatron combination of measurements of the \ttbar production cross section and its differential measurement, the first evidence for and observation of the production of single top quarks in the $s$-channel, as well the final Tevatron combination of the production of single top quarks the $s$- and $t$-channels. Furthermore, I will review the measurements of the forward-backward asymmetry in \ttbar events, which can be experimentally uniquely accessed in the $CP$-invariant $p\bar p$ initial state at the Tevatron, and conclude with the measurements of this asymmetry in the $b\bar b$ system.
}

The pair-production of the top quark was discovered in 1995 by the CDF and D0 experiments~\cite{bib:topdiscovery} at the Fermilab Tevatron proton-antiproton collider. Observation of the electroweak production of single top quarks was presented only six years ago~\cite{bib:singletop}. The large top quark mass~\cite{bib:mt_tev}
indicates that the top quark could play a crucial role in electroweak symmetry breaking. Precise measurements of the production of the top quark and of its properties provide a crucial test of the consistency of the SM and could hint at physics beyond the standard model (BSM). This article discusses recent precision measurements of the top production at the Tevatron, while recent measurements of the intrinsic properties of the top quark, including the world's most precise single measurement of the top quark mass \mt~\cite{bib:mt}, are covered in Ref.~\cite{bib:prop}. All Tevatron results in the top sector can be found in~Refs.~\cite{bib:toprescdf,bib:topresd0}. 
The corresponding measurements by the LHC experiments are discussed in Refs.~\cite{bib:lhctop}.

Top quarks are mostly produced in pairs via the strong interaction. At the Tevatron, the production process $q\bar q\to\ttbar$ dominates with a contribution of $\approx85\%$ over the $gg\to\ttbar$ reaction, while at the LHC their respective fractions are approximately reversed. Another striking difference between the two colliders is that the $p\bar p$ initial state at the Tevatron is an eigenstate of the $CP$ transformation. The relative contribution of the $s$-channel to the total production cross section of single top quarks $\sigma_t$ is $\approx30\%$ at the Tevatron, making it experimentally easier to access than at the LHC, where it contributes only $\approx5\%$.
In the SM, the top quark decays to a $W$ boson and a $b$ quark nearly 100\% of the time. Thus, $\ttbar$ events decay to a $W^+W^-b\bar b$ final state, which is further classified according to the $W$ boson decays into ``dileptonic'', ``\ljets'', or ``all--jets'' channels. Accordingly, single top production in the $s$-channel proceeds dominantly through $q\bar q\to t\bar b$, while the $qg\to q't\bar b$ process dominates in the $t$-channel.

\section{Production of $\boldsymbol{\ttbar}$ pairs} 

\begin{figure}
\centering
\begin{overpic}[clip,height=4.5cm]{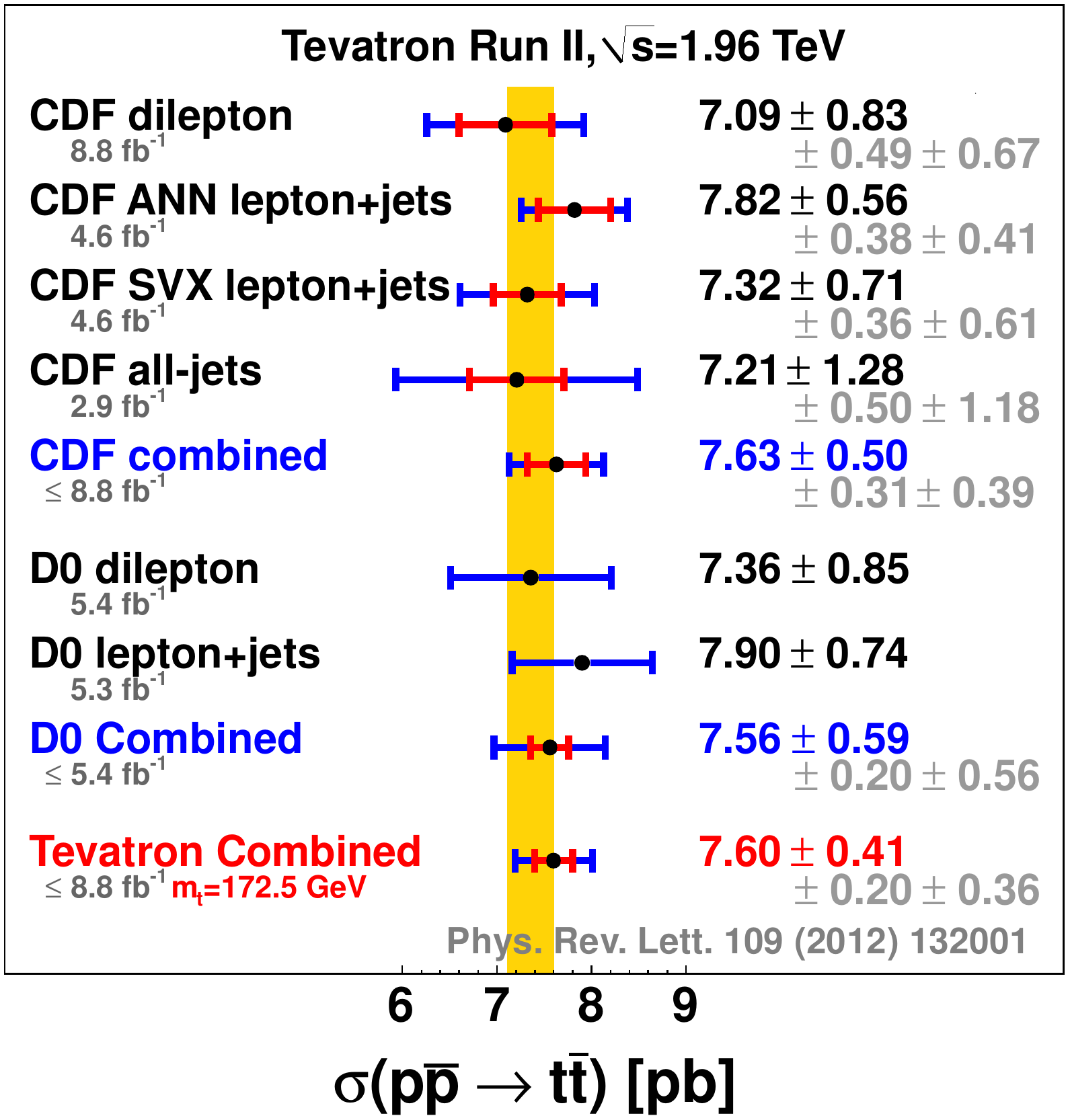}
\put(-1,2){(a)}
\end{overpic}
\begin{overpic}[clip,height=4.5cm]{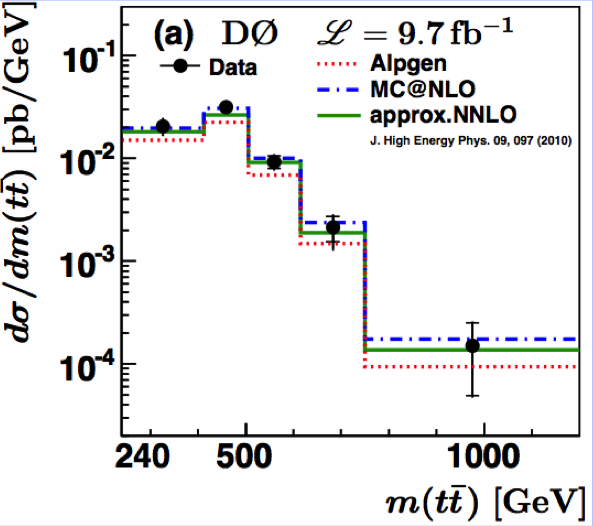}
\put(-1,2){(b)}
\end{overpic}
\begin{overpic}[clip,height=4.5cm]{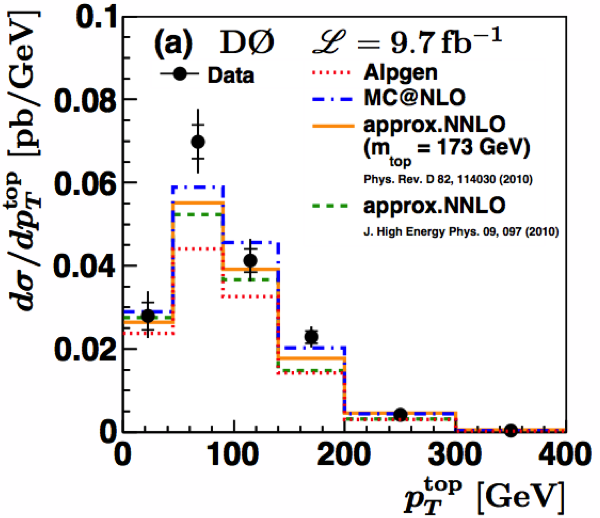}
\put(-1,2){(c)}
\end{overpic}
\caption{\label{fig:xsec_tt}
{\bf(a)} The combination of $\stt$ from CDF and D0 in $p\bar p$ collisions at $\sqrt s=1.96~\TeV$ compared to the SM prediction at NNLO+NNLL. 
{\bf(b)} The distribution of $\dif\stt/\dif m_{\ttbar}$ after background subtraction and regularised unfolding measured by D0 in the \ljets channel using 9.7~\fb of data, compared with theory predictions.
{\bf(c)} Same as (b), but for $\dif\stt/\dif \pt^{\rm top}$. Both $t$ and $\bar t$ contribute in each event in~(c).
}
\end{figure}

The CDF and D0 Collaborations combined their most precise measurements of the \ttbar production cross section \stt in Run II of the Tevatron~\cite{bib:xsec_tt}, summarised in Fig.~\ref{fig:xsec_tt}~(a). The combination is performed using the best linear unbiased estimator (BLUE) technique, considering all sources of systematic uncertainties and their correlations.
The combined value of $\stt=7.60\pm0.41$~pb is in good agreement with the SM prediction of $\stt=7.35^{+0.28}_{-0.33}$~pb~\cite{bib:xsec_tt_nnlo}, which is calculated at next-to-next-to-leading order (NNLO) with next-to-next-to-leading logarithmic (NNLL) corrections.

D0 recently measured $\stt$ differentially in various kinematic distributions in the \ljets chanel using 9.7~\fb of data~\cite{bib:xsec_tt_diff}. After subtraction of backgrounds, the distributions are corrected for detector effects through a regularised matrix unfolding. A representative selection of results is shown in Fig.~\ref{fig:xsec_tt}~(b) for $\dif\stt/\dif m_{\ttbar}$ and in~(c) for $\dif\stt/\dif \pt^{\rm top}$. The SM MC simulations considered are able to describe the differential dependence up to an overall normalisation factor.

\section{Production of single top quarks}

\begin{figure}
\centering
\begin{overpic}[clip,height=5cm]{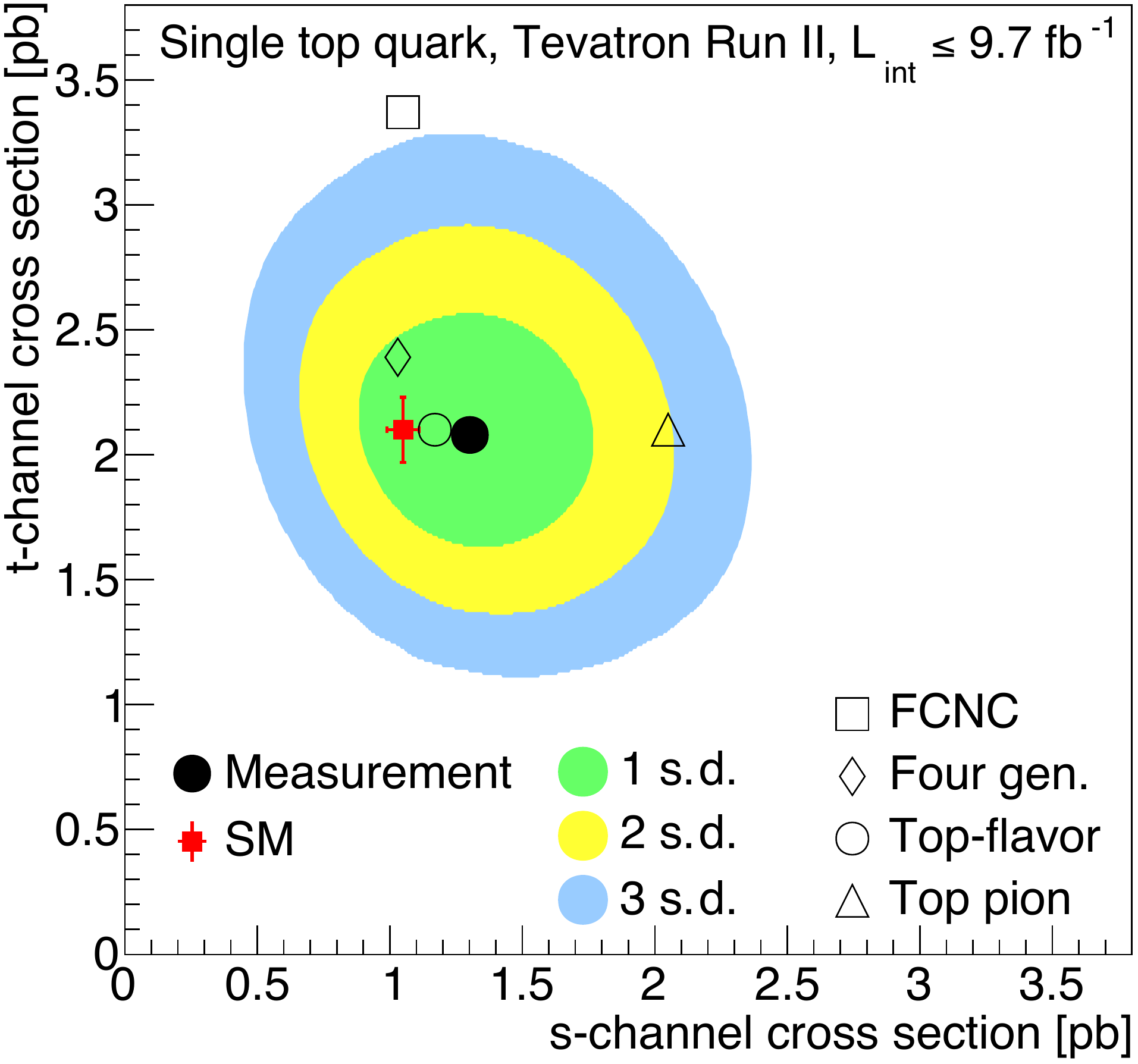}
\put(-1,2){(a)}
\end{overpic}
\begin{overpic}[clip,height=5cm]{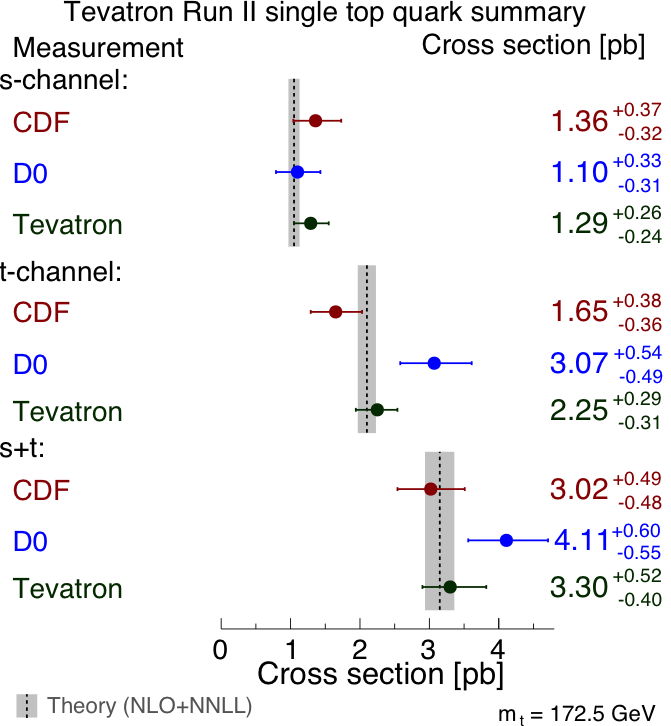}
\put(-1,7){(b)}
\end{overpic}
\caption{\label{fig:st}
{\bf(a)} The combination of measured cross sections for the production of single top quarks in the $s$- and $t$-channel compared to SM expectation and BSM predictions. 
{\bf(b)} The overview of the measurements of single top quark production in the $s$- and $t$-channels.
}
\end{figure}

The production of single top quarks in the $s$-channel is difficult to observe at the LHC given its small cross section of $\sigma_{s-{\rm ch.}}=5.5\pm0.2$~pb~\cite{bib:kido}. The experimental situation is less challenging at the Tevatron given that $\sigma_{s-{\rm ch.}}=1.05\pm0.06$~pb contributes about 1/3 of the total $\sigma_t$.

The first evidence for single top production was provided by D0~\cite{bib:sch_ev_d0} in the \ljets channel using 9.7~\fb of data. The sensitivity of the analysis was enhanced by categorising events into signal-enriched and signal-depleted categories according to the jet and $b$-tag multiplicity.
Two discriminants sensitive to the $s$- and $t$-channel were used to extract $\sigma_{s-{\rm ch.}}$ and $\sigma_{t-{\rm ch.}}$ simultaneously. Without any assumption on $\sigma_{t-{\rm ch.}}$, D0 measures $\sigma_{s-{\rm ch.}}=1.10\pm0.33$~pb, which corresponds to a significance of 3.7 standard deviations (SD). Recently, CDF also reported an evidence for single top production in the $s$-channel at the level of 4.2~SDs~\cite{bib:sch_ev_cdf_lj} using 9.5~\fb of data. They combined the analyses in two channels, \ljets~\cite{bib:sch_ev_cdf_lj} and $\met\!+\!{\rm jets}$ (where the lepton is missed)~\cite{bib:sch_ev_cdf_mj}, both using a discriminant optimised to the $s$-channel only. By combining these three measurements, CDF and D0 discovered single top production in the $s$-channel with an observed (expected) significance of 6.3 (5.1)~SDs~\cite{bib:sch_obs}.

CDF and D0 performed the final combination of their best measurements in the $s$- and $t$ channels by constructing a joint discriminant and considering all uncertainties and their correlations~\cite{bib:st_combi}. The simultaneously extracted $\sigma_{s-{\rm ch.}}=1.29\pm0.25$~pb and $\sigma_{t-{\rm ch.}}=2.25\pm0.30$~pb, shown in Fig.~\ref{fig:st}~(a), agree with the SM expectation and, combined, can exclude a number of BSM scenarios. The overview of single top measurements at the Tevatron is given in Fig.~\ref{fig:st}~(b).

\section{Forward-backward asymmetry in $\boldsymbol{t\bar t}$ events}

\begin{figure}
\begin{overpic}[clip,height=3.7cm]{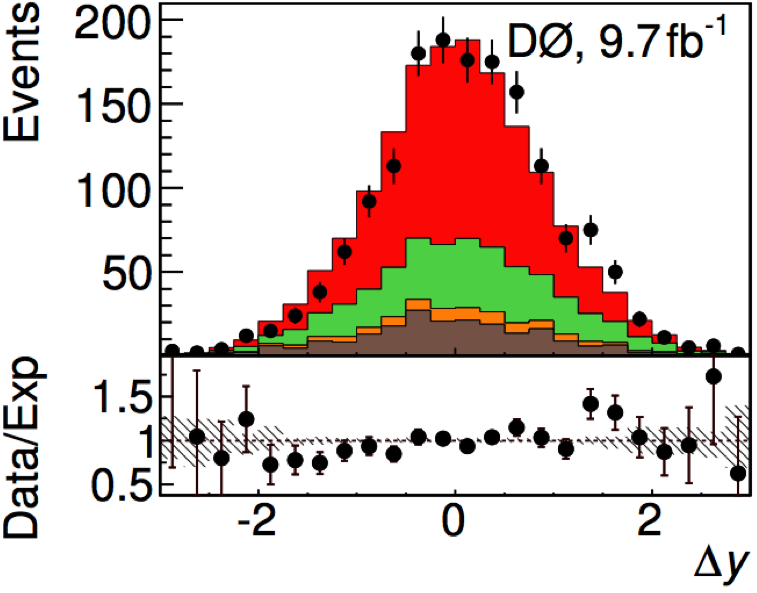}
\put(5,2){(a)}
\end{overpic}
\begin{overpic}[clip,height=3.7cm]{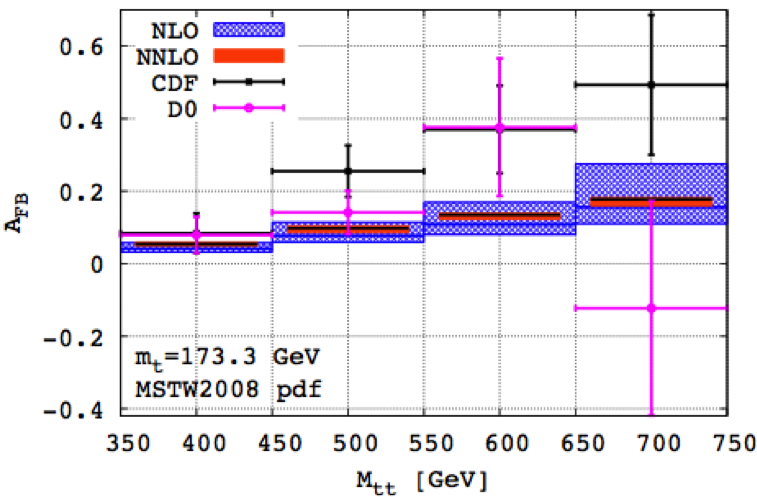}
\put(-1,2){(b)}
\end{overpic}
\begin{overpic}[clip,height=2.3cm]{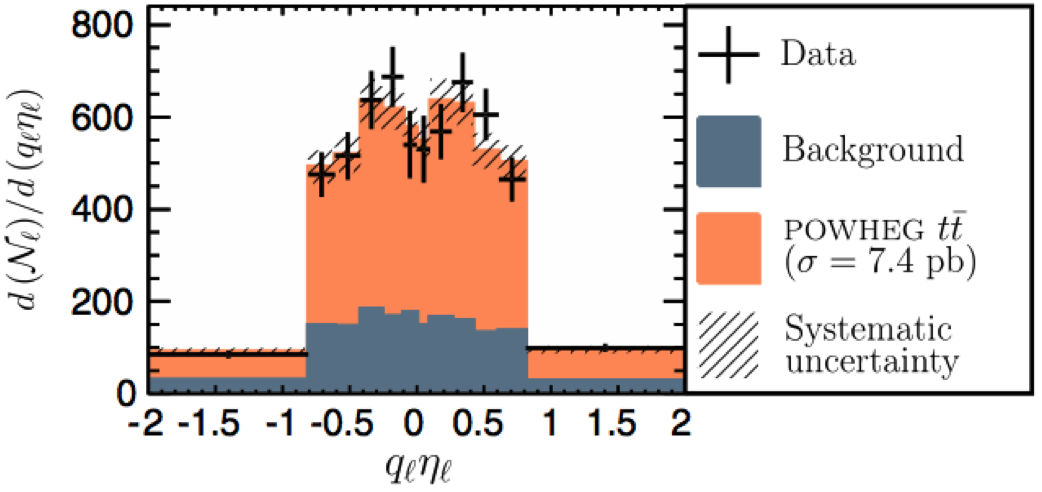}
\put(-1,2){(c)}
\end{overpic}
\caption{
\label{fig:afb_meas}
{\bf(a)} The distribution in $\Delta y$ in data compared to simulations in the \ljets channel using 9.7~\fb of D0 data.
The data prefers a larger asymmetry than MC@NLO that is used to simulate \ttbar events.
{\bf(b)} The distribution in $\afbtt$ in the \ttbar\ rest frame versus $M_{\ttbar}$, after background subtraction and corrected for experimental effects from the CDF
and D0
experiments, is compared to NNLO+NNLL calculations.
{\bf(c)} The distribution in $q\eta_\ell$ in data compared with sumulations in the dilepton channel using 9.1~\fb of CDF data.
}
\end{figure}

In the SM, the pair production of top quarks in $p\bar p$ collisions exhibits a forward-backward asymmetry $\afbtt\equiv\frac{N^{\Delta y>0}-N^{\Delta y<0}}{N^{\Delta y>0}+N^{\Delta y>0}}$ of $\approx10\%$ at NNLO with NNLL corrections in the $\ttbar$ rest frame~\cite{bib:afbtt_nnlo}, where $\Delta y\equiv y_t-y_{\bar t}$ and $y_t=\frac12\ln\frac{E_t+p_{z,t}}{E_t-p_{z,t}}$ is the rapidity of the $t$ quark. 

\enlargethispage{0.5cm}

D0 measured $\afbtt=10.6\pm3.0\%$ in the \ljets channel using events with at least one $b$-tagged jet in 9.7~\fb of data~\cite{bib:afbtt_d0_lj}, as shown in Fig.~\ref{fig:afb_meas}~(a). The kinematic fitter used for the reconstruction was extended to cover events with only three jets. 
A similar measurement was performed by CDF in  the \ljets channel~\cite{bib:afbtt_cdf_lj} using 9.4~\fb of data. Their result of $\afbtt=16.4\pm4.7\%$ is $\approx1.5$~SD away from the SM expectation. 
Both Collaborations also investigated the dependence of $\afbtt$ on the invariant mass of the $\ttbar$ system $m_{\ttbar}$ which is compared to SM expectation~\cite{bib:afbtt_nnlo} in Fig.~\ref{fig:afb_meas}~(b), and on $|\Delta y|$. 

The lepton-based asymmetry $\afbl\equiv\frac{N^{q\eta>0}-N^{q\eta<0}}{N^{q\eta>0}+N^{q\eta>0}}$ is sensitive to $\afbtt$ through the charge-signed pseudorapidity $q\eta$ of charged leptons in $t\to\ell\nu b$ decays. CDF measured $\afbl=7.2\pm6.0\%$ in the dilepton channel using 9.1~\fb of data, as shown in Fig.~\ref{fig:afb_meas}~(c), while D0 found $\afbl=4.4\pm3.9\%$ using 9.7~\fb. Both results are consistent with SM expectation.

\begin{figure}
\centering
\begin{overpic}[clip,height=6cm]{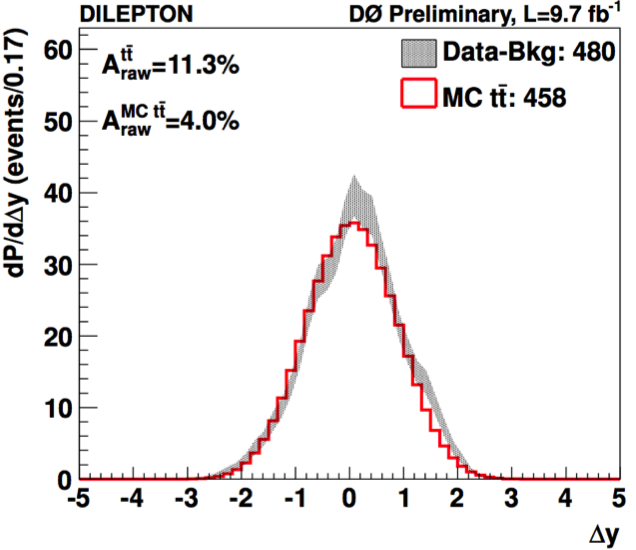}
\put(-1,2){(a)}
\end{overpic}
\begin{overpic}[clip,height=6cm]{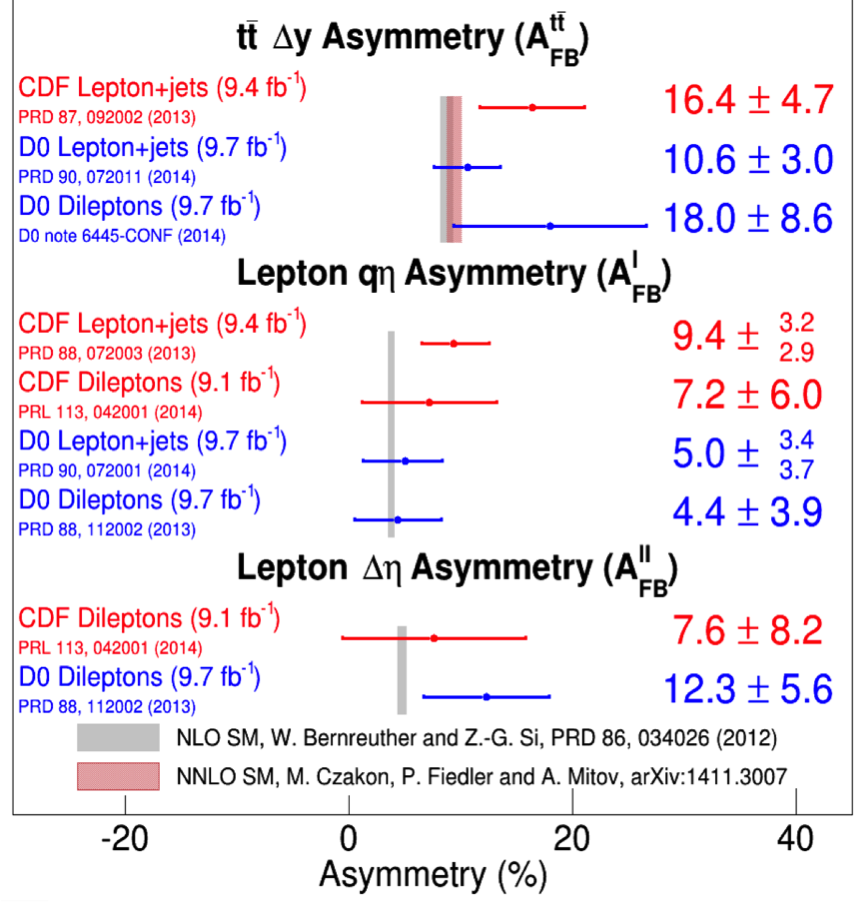}
\put(-1,2){(b)}
\end{overpic}
\caption{\label{fig:asymm}
{\bf(a)} The distribution in $\Delta y$ in data compared to simulations after reconstruction with the ME method and before calibration in the dilepton channel using 9.7~\fb of D0 data.
{\bf(b)} The overview of the measurements of \afbtt, \afbl, and $\afb^{\ell\ell}$~(not discussed here) by CDF and D0, compared to theory predictions. 
}
\end{figure}

D0 applied the matrix element method, which calculates the probability of each event to come from \ttbar production as a function of $\Delta y$, to reconstruct \afbtt in kinematically underconstrained dilepton events~\cite{bib:afbtt_d0_ll}, and found $\afbtt=18.0\pm8.6\%$. The distribution in $\Delta y$ is shown in Fig.~\ref{fig:asymm}~(a).

An overview of all measurements of the forward-backward asymmetry is given in Fig.~\ref{fig:asymm}~(b). 
Both Collaborations have investigated $\afb$ in the $b\bar b$ system at high $M_{b\bar b}$, where $q\bar q\to b\bar b$ dominates~\cite{bib:afb_bb_cdf_hi}. These results are discussed in detail in Ref.~\cite{bib:afb_bb}.

\section*{Acknowledgments}
I would like to thank my collaborators from the CDF and D0 experiments for their help in preparing this article. I also thank the staffs at Fermilab and collaborating institutions, as well as the CDF and D0 funding agencies.

\end{document}